\documentclass[pra,twocolumn,amsmath,amssymb,showpacs]{revtex4}

\usepackage{graphicx}
\usepackage{amssymb}
\usepackage{epstopdf}
\usepackage{times}
\usepackage{dcolumn}
\usepackage{bm}
\usepackage{amsmath}
\usepackage[dvips]{color}

\newcommand{\caA}{\mathcal{A}}
\newcommand{\caB}{\mathcal{B}}

\newcommand{\caN}{\mathcal{N}}

\newcommand{\caS}{\mathcal{S}}
\newcommand{\caR}{\mathcal{R}}

\newcommand{\spaEq}{\hspace{2em}} 
\newcommand{\spaNeg}{\hspace{-2em}} 

\newcommand{\vspfigA}{\vspace{0cm}}  
\newcommand{\vspfigB}{\vspace{0cm}} 
\newcommand{\vspfigC}{\vspace{0.2cm}}
\newcommand{\widthfig}{0.45\textwidth}

\begin{document}


\title{Quantum particle escape from a time-dependent confining potential}

\author{Tooru Taniguchi and Shin-ichi Sawada}

\affiliation{School of Science and Technology, Kwansei Gakuin University, 2-1 Gakuen, Sanda city, Japan} 

\date{\today}

\begin{abstract}

   Quantum escape of a particle via a time-dependent confining potential in a semi-infinite one-dimensional space is discussed. We describe the time-evolution of escape states in terms of scattering states of the quantum open system, and calculate the probability $P(t)$ for a particle to remain in the confined region at time $t$ in the case of a delta-function potential with a time-oscillating magnitude. The probability $P(t)$ decays exponentially in time at early times, then decays as a power later, along with a time-oscillation in itself. We show that a larger time-oscillation amplitude of the confining potential leads to a faster exponential decay of the probability $P(t)$, while it can rather enhance the probability $P(t)$ decaying as a power. These contrastive behaviors of the probability $P(t)$ in different types of decay are discussed quantitatively by using the decay time and the power decay magnitude of the probability $P(t)$. 
   
\end{abstract}


\maketitle 
\section{Introduction}

   Quantum escape, as a quantum mechanical leakage of materials from a confined area, is an important concept to study dynamical properties of quantum open systems. 
   It appears in many physical phenomena, such as the $\alpha$ decay of a nucleus \cite{G28,GC29}, radiative decay of molecules \cite{K10}, transition of states in chemical reactions \cite{T07}, etc. 
   Studies on quantum escapes have been done in one-dimensional systems with a stationary potential barrier \cite{W61,DN02,GMV07}, billiard systems with leads \cite{ZB03}, kicked rotator models \cite{CM97,F97,SS97}, network systems \cite{BG01,PSK05,TS11b}, and many-particle systems \cite{CDG06,TS11a,C11}, and so on. 
   
   In escape phenomena, materials continue escaping from a confined area as an irreversible process, so various quantities decay in time. 
   As one of such decaying quantities, the probability for particles to remain in the confined area, which we call the survival probability \cite{ZB03,PSK05,TS11a,W12,BD07} in this paper, has been widely investigated in escape phenomena.  
   In quantum escapes, the survival probability decays exponentially in time \cite{GMV07,PSK05,W12,M85}, by tunneling via a potential barrier, etc. 
   Moreover, it also shows a power decay after a long time \cite{DN02,ZB03,TS11b,OK91,DHM92}. 
   As decay properties of the survival probability, time scales of its exponential decay \cite{CM97,F97,SS97}, and changes of the power of decay under different conditions  \cite{TS11b,TS11a,C11,M03}, etc. have been studied. 
   
   The principal aim of this paper is to investigate effects of a time-dependent confining potential in quantum escape phenomena. 
   Quantum dynamics of systems with a time-periodic potential is analyzed by the Floquet theory, etc. \cite{S65,R04}, and recent experimental and theoretical developments in microscopic or mesoscopic systems have pushed to study quantum dynamical behaviors caused by a time-dependent external manipulation \cite{T07,R03,NM09,M12}. 
   In this paper, we consider quantum escapes of a particle  via a time-periodic confining potential in a semi-infinite one-dimensional space, as a model of a time-dependent manipulation of particle escapes.  
   In the case of a time-dependent delta-function potential, we obtain an analytical form of the quantum propagator to describe a time-evolution of the wave function, leading to a calculation of the survival probability of the system at any time. 
   We show that a time-oscillation of the confining potential has opposite effects in different decay behaviors of the survival probability.  
   In the exponential decay of the survival probability, a larger time-oscillation amplitude of the confining potential leads to a faster decay of the survival probability.  
   In contrast, the time-oscillation of the confining potential can rather enhance the survival probability decaying as a power. 
   We calculate the decay rate of the survival probability and the magnitude of its power decay, by which these reduction and enhancement of the survival probability by a time-oscillation of the confining potential are discussed quantitatively.

%
   

\section{Quantum open systems with a time-dependent potential}

\subsection{Semi-infinite one-dimensional systems with a time-periodical potential}

   The system, which we consider in this paper, is a quantum particle system with a time-dependent potential in the semi-infinite one-dimensional space. 
   The wave function $\Psi(x,t)$ of the system at position $x$ and time $t$ satisfies the Schr\"udingier equation $i\hbar\partial \Psi(x,t)/\partial t = \hat{H}(t)\Psi(x,t)$. 
   Here, $\hat{H}(t) \equiv - [\hbar^{2}/(2m)]\partial^2/\partial x^{2} + U(x,t)$ is the Hamiltonian operator with the mass $m$ of particle, Planck's constant $2\pi\hbar$, and the potential $U(x,t)$ at position $x$ and time $t$. 
   We impose the hard-wall boundary condition at the end $x=0$ of the semi-infinite one-dimensional space $x \geq 0$, so that the wave function $\Psi(x,t)$ satisfies the condition $\Psi(0,t)=0$ at any time $t$. 
   We assume that the potential $U(x,t)$ is a periodic function of time, so that 
\begin{eqnarray}
   U(x,t) = U(x,t+T) = \sum_{n=-\infty}^{+\infty} u^{(n)}(x) e^{i n \omega t}
\label{TDepenPoten1}
\end{eqnarray}
with the time period $T = 2\pi/\omega$, in which $\omega$ is a positive constant and $u^{(n)}(x)$ is a function of position $x$ only ($n=0,\pm 1,\pm 2,\cdots$). 
   We impose the condition $u^{(-n)}(x) = \left[u^{(n)}(x)\right]^{*}$ for the function $u^{(n)}(x)$ of $x$ in Eq. (\ref{TDepenPoten1}), so that the potential $U(x,t)$ is a real function of $x$ and $t$. 
   Here, $X^{*}$ with the asterisk ${}^{*}$ denotes the complex conjugate of $X$ for any complex number $X$. 

   We consider the scattering state $\Phi_{k}(x,t)$ at position $x$ and time $t$, which is induced by the incident plain wave from the infinite region $x\rightarrow +\infty$ with the energy $E_{k} = (\hbar k)^2/(2m)$ and the wave number $k$. 
   As a solution of the Schr\"udingier equation with the time-dependent potential (\ref{TDepenPoten1}), the scattering state is represented as 
\begin{eqnarray}
   \Phi_{k}(x,t) = \sum_{n=-\infty}^{+\infty} \phi_{k}^{(n)}(x) e^{-i(E_{k}+n\hbar \omega)t/\hbar} ,
\label{ScattState1}
\end{eqnarray}
where the function $\phi_{k}^{(n)}(x)$ of $x$ satisfies the time-independent differential equation \cite{S65,R03} as
\begin{eqnarray}
   && \frac{d^{2} \phi_{k}^{(n)}(x)}{dx^{2}} + \Upsilon_{k}^{(n)} \phi_{k}^{(n)}(x) 
      \nonumber \\
   && \spaEq 
      = \frac{2m}{\hbar^2}\sum_{\mu =-\infty}^{+\infty} u^{(\mu-n)}(x)  \phi_{k}^{(\mu)}(x) 
\label{SchruEquat1}
\end{eqnarray}
with $\Upsilon_{k}^{(n)}$ defined by
\begin{eqnarray}
   \Upsilon_{k}^{(n)} \equiv  k^{2}+\frac{2m\omega}{\hbar} n .
\label{Upsi1}
\end{eqnarray}
   From Eq. (\ref{SchruEquat1}) we derive the Lippmann-Schwinger equation 
\begin{widetext}
\begin{eqnarray}
   \phi_{k}^{(n)}(x) = \delta_{n0} \sqrt{\frac{2}{\pi}}\sin(kx) 
   + \frac{2m}{\hbar^2}\sum_{\mu =-\infty}^{+\infty} 
      \int_{0}^{+\infty} dy\; G_{k}^{(n)}(x,y) 
      u^{(\mu-n)}(y)  \phi_{k}^{(\mu)}(y)
\label{LippmSchwi1}
\end{eqnarray}
\end{widetext}
where $G_{k}^{(n)}(x,y)$ is given by
\begin{eqnarray}
   G_{k}^{(n)}(x,y) \equiv 
   \frac{e^{i\sqrt{\Upsilon_{k}^{(n)}}|x-y|} 
      -e^{i\sqrt{\Upsilon_{k}^{(n)}}|x+y|}}{2i\sqrt{\Upsilon_{k}^{(n)}}} ,
\label{GreenFunct1}
\end{eqnarray}
as a Green function satisfying the differential equation 
\begin{eqnarray}
   \frac{\partial^{2} G_{k}^{(n)}(x,y)}{\partial x^{2}} + \Upsilon_{k}^{(n)} G_{k}^{(n)}(x,y)
   = \delta (x-y)
\label{GreenDiffe1}
\end{eqnarray}
with the boundary condition $G_{k}^{(n)}(0,y) = G_{k}^{(n)}(x,0) =0$. 
   Here, in the scattering state (\ref{ScattState1}) we imposed the condition that if there is no potential energy then the scattering state $\Phi_{k}(x,t)$ is represented as the state $\psi_{k}(x,t) \equiv \sqrt{2/\pi}\sin(kx)\exp(-iE_{k}t/\hbar)$ induced by the incident plain wave with the wave number $k$, with the orthogonal relation $\int_{0}^{+\infty} dx \; [\psi_{k}(x,t)]^{*}\psi_{k'}(x,t)= \delta(k-k')$ for $kk'>0$ and the boundary condo $\psi_{k}(0,t)$=0. 
   This condition determines the form $\delta_{n0} \sqrt{2/\pi}\sin(kx)$ of the first term in the right-hand side of Eq. (\ref{LippmSchwi1}). 
   We also defined the square root of $\Upsilon_{k}^{(n)}$ as $\sqrt{\Upsilon_{k}^{(n)}} \equiv i\sqrt{|\Upsilon_{k}^{(n)}|}$ for the case of $\Upsilon_{k}^{(n)}<0$ in Eq. (\ref{GreenFunct1}).
   More explanations of Eq. (\ref{LippmSchwi1}) and the Green function (\ref{GreenFunct1}) are given in Appendix \ref{GreenFunctionHardWall}.

\subsection{Scattering states with a time-dependent delta-function potential}
   
   From now on we especially consider the case of 
\begin{eqnarray}
   u^{(n)}(x) = \lambda^{(n)}\delta (x-L)
\label{DeltaPoten1}
\end{eqnarray}
with constants  $L \; (>0)$ and $\lambda^{(n)}$, satisfying the condition $\lambda^{(-n)} = \left[\lambda^{(n)}\right]^{*}$, $n=0,\pm 1,\pm 2,\cdots$. 
   Under the condition (\ref{DeltaPoten1}) the potential is represented as $U(x,t) = \delta (x-L) \sum_{n=-\infty}^{+\infty} \lambda^{(n)} e^{i n \omega t}$, i.e. the delta-function potential with a time-dependent magnitude. 
   By Eq. (\ref{DeltaPoten1}), Eq. (\ref{LippmSchwi1}) is transformed into
\begin{eqnarray}
   \phi_{k}^{(n)}(x) &=& \delta_{n0} \sqrt{\frac{2}{\pi}}\sin(kx) 
      \nonumber \\
   && 
      + \frac{2m}{\hbar^2}
      G_{k}^{(n)}(x,L) \sum_{\mu =-\infty}^{+\infty} \lambda^{(\mu-n)}  \phi_{k}^{(\mu)}(L)
\label{LippmSchwi2}
\end{eqnarray}
without any spatial integration. 

   We introduce the matrix $\Lambda_{k} = (\Lambda_{k}^{(\mu\nu)})$ with the $\mu\nu$ matrix element 
\begin{eqnarray}
   \Lambda_{k}^{(\mu\nu)}\equiv \frac{2m}{\hbar^{2}} G_{k}^{(\mu)}(L,L) \lambda^{(\nu-\mu)} .
\label{FunctLambda1}
\end{eqnarray}
   Using the matrix element (\ref{FunctLambda1}), Eq.  (\ref{LippmSchwi2}) at $x=L$ leads to the relation $\sum_{\mu=-\infty}^{+\infty} \left[\delta_{n\mu}-\Lambda_{k}^{(n\mu)}\right]\phi_{k}^{(\mu)}(L) = \delta_{n0} \sqrt{\frac{2}{\pi}}\sin(k L)$, namely  
\begin{eqnarray}
   \phi_{k}^{(n)}(L) 
   &=&  \sqrt{\frac{2}{\pi}}\sin(kL) \left[(I-\Lambda_{k})^{-1} \right]^{(n0)} 
\label{ValuePhiL1}
\end{eqnarray}
with the identical matrix $I$. 
   Here and hereafter we put $X^{(\mu\nu)}$ as the $\mu\nu$ element of $X$ for any matrix $X$ in this paper. 
   By inserting Eq. (\ref{ValuePhiL1}) into Eq. (\ref{LippmSchwi2}) and using Eq. (\ref{FunctLambda1}) we obtain the quantity $\phi_{k}^{(n)}(x)$ as 
\begin{eqnarray}
   \phi_{k}^{(n)}(x) 
   &=& \delta_{n0} \sqrt{\frac{2}{\pi}}\sin(kx) 
      \nonumber \\
      &&
   + \sqrt{\frac{2}{\pi}} \frac{G_{k}^{(n)}(x,L)\sin(kL)}{G_{k}^{(n)}(L,L) }
   \left[\Lambda_{k}(I-\Lambda_{k})^{-1} \right]^{(n0)} .
      \nonumber \\
      &&
\label{LippmSchwi3}
\end{eqnarray}
The scattering state $\Phi_{k}(x,t)$ induced by the incident plain wave with the wave number $k$ is given by Eq. (\ref{ScattState1}) with Eq. (\ref{LippmSchwi3}). 

   It is essential to note for contents of this paper that the scattering state $\Phi_{k}(x,t)$ satisfies the orthogonal relation as 
\begin{eqnarray}
   \int_{0}^{+\infty} dx \; \left[\Phi_{k}(x,t)\right]^{*}\Phi_{k'}(x,t)  = \delta(k-k') 
\label{OrthoRelat1}
\end{eqnarray}
for any time $t$ and almost all positive numbers $k$ and $k^{'}$. 
   The proof and explanations of Eq. (\ref{OrthoRelat1}) are given in Appendix \ref{OrthogonalRelation}.

\subsection{Time-evolution of the wave function with the propagator}

   In this paper we restrict ourselves our considerations into the quantum state $\Psi (x,t)$ which is expanded by the scattering state $\Phi_{k}(x,t)$ as
\begin{eqnarray}
   \Psi (x,t) = \int_{0}^{+\infty}dk\; A_{k}\Phi_{k}(x,t)
\label{WaveFunct1}
\end{eqnarray}
with the coefficient $A_{k}$. 
   By using Eqs. (\ref{OrthoRelat1}) and (\ref{WaveFunct1}), the inner product $\int_{0}^{+\infty} dx \; \left[\Phi_{k}(x,0)\right]^{*}\Psi(x,0) $ of $\Phi_{k}(x,0)$ and $\Psi (x,0)$ leads to the coefficient $A_{k}$, namely 
\begin{eqnarray}
   A_{k} = \int_{0}^{+\infty} dx \; \left[\Phi_{k}(x,0)\right]^{*}\Psi(x,0) .
\label{CoeffA1}
\end{eqnarray}
   By inserting Eq, (\ref{CoeffA1}) into Eq. (\ref{WaveFunct1}), the wave function $\Psi (x,t)$ at position $x$ and time $t$ is rewritten as 
\begin{eqnarray}
   \Psi (x,t) = \int_{0}^{+\infty} dy\; K(x,y;t)  \Psi (y,0) , 
\label{WaveFunct2}
\end{eqnarray}
where $K(x,y;t)$ is defined by 
\begin{eqnarray}
   K(x,y;t) \equiv \int_{0}^{+\infty} dk \; \Phi_{k}(y,0)^{*} \Phi_{k}(x,t) 
\end{eqnarray}
as the propagator for the time-evolution of the wave function $\Psi (x,t)$.

\section{Particle escape from a time-oscillating confining potential}

\subsection{Survival probability} 

   Now we consider the initial condition in which the particle exists only in the finite region $(0,L)$  at the initial time $t=0$, namely   
\begin{eqnarray}
   \Psi (x,0) =    \left\{\begin{array}{ll}
     \Psi_{0} (x) & \mbox{for} \;\;\; 0\leq x < L   \\
       0 &\mbox{for} \;\;\;  L \leq x  
   \end{array}\right. 
\label{InitiCondi1}
\end{eqnarray}
with a function $\Psi_{0} (x)$ of $x$. 
   Staring from this initial condition, we investigate the probability $P(t)$ for a particle in the finite region $(0,L)$ at time $t$ as 
\begin{eqnarray}
   P(t) \equiv \int_{0}^{L}dx\; \left|\Psi (x,t) \right|^{2} , 
\label{SurviProba1}
\end{eqnarray}
which we call the survival probability.

  Furthermore, from now on we assume that the initial state is given by the energy eigenstate of the particle confined in the finite region $[0,L]$, namely, 
\begin{eqnarray}
   \Psi_{0}(x) = \sqrt{\frac{2}{L}}\sin\!\left(\frac{\sigma\pi}{L}x\right) 
\label{InitiCondi2}
\end{eqnarray}
for $0\leq x \leq L$ with an integer value $\sigma = 1,2,\cdots$. 
   Under this specification of the initial conditions (\ref{InitiCondi1}) and (\ref{InitiCondi2}) we can carry out analytically the integral over the position $x$ in Eq. (\ref{SurviProba1}). 
   We give an explicit form of the survival probability $P(t)$ in Appendix \ref{SurvivalProbability} under this initial condition. 

   As a concrete time-dependence in the potential $U(x,t)$, in the next subsection \ref{TimeDependentPotentialEffects} we consider a time-oscillating potential, i.e. 
\begin{eqnarray}
   U(x,t) = [a + b\cos(\omega t)]\delta(x-L)
\label{TDepenPoten2}
\end{eqnarray}
with real constants $a$ and $b$. 
   This potential is derived from 
\begin{eqnarray}
   \lambda^{(n)} = \left\{\begin{array}{ll}
      a           &  \mbox{for} \;\;\; n=0  \\
      \frac{b}{2} & \mbox{for} \;\;\; n=\pm 1     \\
      0           & \mbox{otherwise}  
   \end{array}\right.
\end{eqnarray}
for the constants $\lambda^{(n)}$ to specify the potential $U(x,t)$ by Eqs. (\ref{TDepenPoten1}) and (\ref{DeltaPoten1}).

\begin{figure}[!t] 
\vspfigA
\begin{center}
\includegraphics[width=\widthfig]{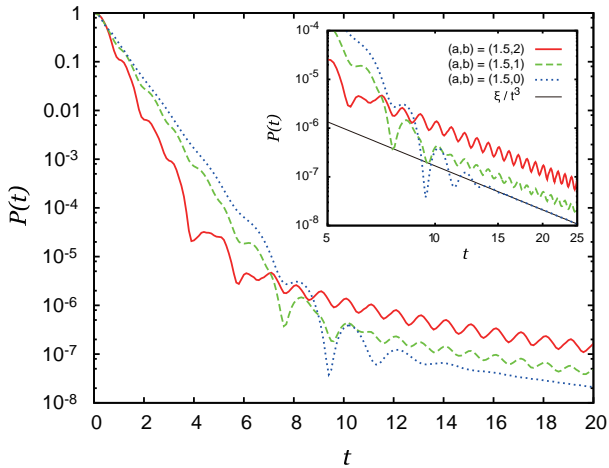}
\vspfigB
\caption{
   (Color online) 
   The survival probabilities $P(t)$ as functions of time $t$ for the cases of $(a,b) = (1.5,2)$ (the solid red line),  $(a,b) = (1.5,1)$ (the dashed green line) and $(a,b) = (1.5,0)$ (the dotted blue line). 
   The main figure is the linear-log plots of $P(t)$, and the inset is their log-log plots as well as the asymptotic power decay $\xi/t^{3}$ corresponding to the case of $(a,b) = (1.5,0)$ shown in the straight shin black line. 
   Here, and in all figures hereafter, we use dimensionless units with $L=1$, $m=1$ and $\hbar =1$.}
\label{Fig2aSurProb1d5}
\end{center}
\vspfigC
\end{figure}  
%
\subsection{Time-oscillating potential effects in decay of the survival probability} 
\label{TimeDependentPotentialEffects} 

   In this subsection we show graphs of the survival probabilities $P(t)$ as functions of time $t$ under the initial conditions (\ref{InitiCondi1}) and (\ref{InitiCondi2}) with the time-dependent potential (\ref{TDepenPoten2}) for various values of the parameters $a$ and $b$. 
   For calculations of such survival probabilities we use the parameter values as $L=1$, $\omega = 2\pi$ [corresponding to the time-period $T =2\pi/\omega =1$ of the potential (\ref{TDepenPoten2})], $\sigma =1$, $m=1$ and $\hbar =1$. 
   In actual numerical calculations we replaced the integral with respect to the wave number $k$ up to $+\infty$ by the integral with respect to the wave number $k$ up to $20$, checking that the absolute value of the actual integrand is extremely small at $k=20$ so that the contribution of the integral over $k>20$ would be negligible. 
   For numerical calculations of the probability $P(t)$ we also have to replace the infinity times infinity matrix $\Lambda_{k}$ with a finite $(2\caN +1)$ times $(2\caN +1)$ matrix $(\Lambda_{k}^{(\mu\nu)})$ with $\mu = -\caN,-\caN+1,\cdots,\caN$ and $\nu = -\caN,-\caN+1,\cdots,\caN$. 
   This implies that we take into account of effects up to the $\caN$-th closest energy scattering states from the state with the wave number $k$ for a small magnitude of the time-depending part of the potential (\ref{TDepenPoten2}).    
   Numerical calculations of the probability $P(t)$ with a large dimensional matrix $(\Lambda_{k}^{(\mu\nu)})$ is very time-consuming, and we used the $7$ times $7$ matrix $(\Lambda_{k}^{(\mu\nu)})$, i.e. $\caN = 3$, for the numerical results in this subsection \ref{TimeDependentPotentialEffects}. 
   For some cases discussed in this subsection we calculated the probabilities $P(t)$ not only for $\caN = 3$ and but also for $\caN = 4$ with the same values of the other parameters. 
   Differences between these calculated results for $\caN = 3$ and $\caN = 4$ were almost negligible comparing to values of the probabilities $P(t)$ themselves, although these results were partly different from the ones for $\caN = 2$. 
   These facts would justify to calculate the probability $P(t)$ by $\caN = 3$ for values of the parameters $a$ and $b$, etc., shown in this subsection \ref{TimeDependentPotentialEffects}.

   Figure \ref{Fig2aSurProb1d5} is the survival probabilities $P(t)$ as functions of time $t$ for the cases of $(a,b) = (1.5,2)$ (the solid red line), $(a,b) = (1.5,1)$ (the dashed green line) and $(a,b) = (1.5,0)$ (the dotted blue line), as their linear-log plots (the main figure) and their log-log plots (the inset) \cite{Memo2a}. 
   Here, in its inset we added the shin black line showing the asymptotic power decay of the survival probability in time for the time-independent potential with $b=0$, as given analytically by $\left.P(t)\right|_{b=0} \stackrel{t\rightarrow+\infty}{\sim} \xi/t^{3}$ with $\xi \equiv 4m^{3}L^{6}\hbar^{5}/[3\sigma^{2}\pi^{3}(\hbar^{2}+2mLa)^{4}]$ \cite{TS11b}. 
   It is shown in this figure that the survival probability $P(t)$ itself oscillates in time for the case of $b\neq 0$ by a time-oscillation of the potential (\ref{TDepenPoten2}). 
   Figure \ref{Fig2aSurProb1d5} also shows that there are two types of decay behaviors of the survival probability $P(t)$; one is an exponential decay at early times, and the other is a power decay later, apart from its time-oscillatory behavior.  
   Moreover, a comparison of the time-independent potential case of $(a,b)=(1.5,0)$ with the cases of $(a,b)=(1.5,1)$ and $(1.5,2)$ in Fig. \ref{Fig2aSurProb1d5} suggests that the time-oscillation of the potential reduces values of the survival probability $P(t)$ decaying exponentially, while it can increase values of the probability $P(t)$ decaying as a power. 
   In the following subsections \ref{ExponentialDecayPart} and \ref{PowerDecayPart} we will discuss these behaviors caused by a time-oscillating potential in different types of decay of the survival probability separately in more details.

\begin{figure}[!t] 
\vspfigA
\begin{center}
\includegraphics[width=\widthfig]{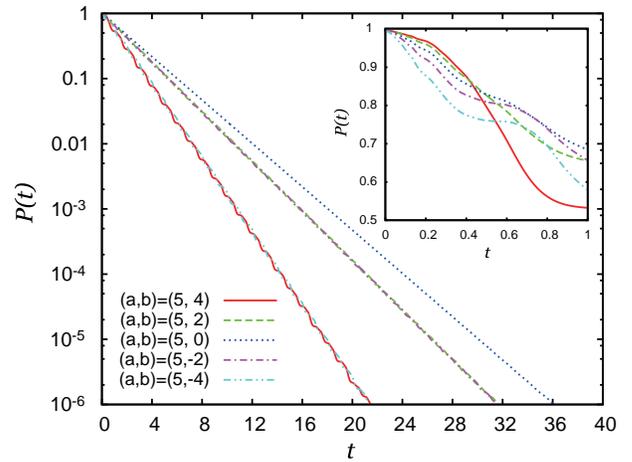}
\vspfigB
\caption{
   (Color online) 
   The survival probabilities $P(t)$ as functions of time $t$ for the cases of $(a,b) = (5,4)$ (the solid red line), $(a,b) = (5,2)$ (the dashed green line), $(a,b) = (5,0)$ (the dotted blue line), $(a,b) = (5,-2)$ (the dash-dotted purple line) and $(a,b) = (5,-4)$ (the dash-double-dotted cyan line). 
   The main figure is the linear-log plots of $P(t)$, and the inset is their linear-linear plots up to the first oscillating period $T=2\pi/\omega = 1$ of the potential. }
\label{Fig2bSurProb5d0}
\end{center}
\vspfigC
\end{figure}  
%
\begin{figure}[!t] 
\vspfigA
\begin{center}
\includegraphics[width=\widthfig]{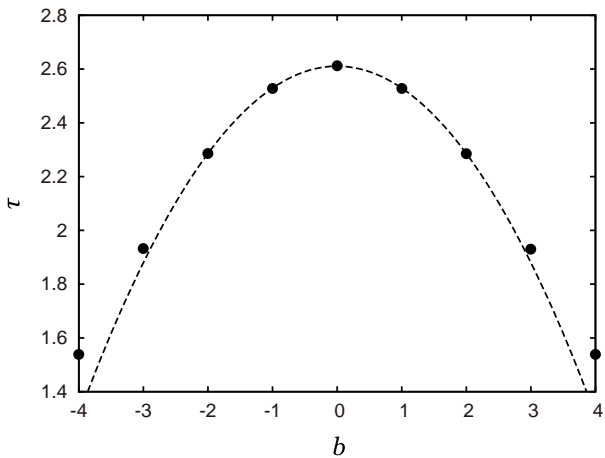}
\vspfigB
\caption{
   The decay time $\tau$ of the survival probability decaying exponentially as a function of the parameter $b$ of the potential $U(x,t) = [a + b\cos(\omega t)]\delta(x-L)$ for the case of $a=5$. 
   The dashed line is a fitting quadratic curve $\tau = \zeta_{1}-\zeta_{2}b^{2}$ for the numerical data of $\tau$ with fitting parameters $\zeta_{1}$ and $\zeta_{2}$. 
   }
\label{Fig2cDecayRate5d0}
\end{center}
\vspfigC
\end{figure}  
%
\subsubsection{Exponential decay of the survival probability} 
\label{ExponentialDecayPart}

   Figure \ref{Fig2bSurProb5d0} is the survival probabilities $P(t)$ as functions of time $t$ for the cases of $(a,b) = (5,4)$ (the solid red line), $(a,b) = (5,2)$ (the dashed green line), $(a,b) = (5,0)$ (the dotted blue line), $(a,b) = (5,-2)$ (the dash-dotted purple line) and $(a,b) = (5,-4)$ (the dash-double-dotted cyan line). 
   Here, the main figure is the linear-log plots of $P(t)$, and the inset is their linear-linear plots at beginning times up to the first oscillating period $T=2\pi/\omega = 1$ of the potential (\ref{TDepenPoten2}). 
   For this figure, we chose a larger value of the parameter $a$ than that in Fig. \ref{Fig2aSurProb1d5}, so that the survival probabilities show their exponential decays more clearly.

   The almost straight lines in the main figure of Fig. \ref{Fig2bSurProb5d0} as linear-log plots of $P(t)$ show that at the presented times the survival probabilities $P(t)$ decay exponentially in time, apart from their small time oscillations with the period $T$.  
   These exponential decays continue for a longer time for a larger value of the parameter $a$ of the potential (\ref{TDepenPoten2}), suggesting that it is caused by a quantum tunneling via the delta-function potential. 
   The survival probability $P(t)$ for $b\neq 0$ oscillates in time with a larger amplitude for a larger time-oscillation amplitude $|b|$ of the potential (\ref{TDepenPoten2}).  
   The survival probabilities with the same absolute values of the parameter $b$, such as $b=\pm 2$ or $b=\pm 4$, decay exponentially in time with the same decay rate but with the opposite phase of their time-oscillations.

   The time-dependent potential barrier magnitude $a + b\cos(\omega t)$ of the potential (\ref{TDepenPoten2}) is larger for a larger value of $b$ for $0 \leq t < T/4$ (and for $3T/4 < t < T$), so that the survival probability $P(t)$ decays slower for a larger value of $b$, as shown in the inset of Fig. \ref{Fig2bSurProb5d0}. 
   After it, for $T/4 < t < 3T/4$ the potential barrier magnitude becomes to be smaller for a larger value of $b$, so that the survival probability $P(t)$ should decay faster for a larger value of $b$ at this time. 
   As shown in the inset of Fig. \ref{Fig2bSurProb5d0}, the magnitude of such an enhancement of the decay speed of the survival probability $P(t)$ in  $T/4 < t < 3T/4$ is larger than that of its suppression in $0 \leq t < T/4$ and $3T/4 < t < T$ for a larger value of $b$. 
   This kind of unbalance of decay speeds of the survival probability $P(t)$ for each oscillating period $T$ of the potential (\ref{TDepenPoten2}) causes a faster decay of the probability $P(t)$ on average for a larger value of the time-oscillation amplitude $|b|$ of the potential (\ref{TDepenPoten2}) as shown in the main figure of Fig. \ref{Fig2bSurProb5d0}.

   In order to discuss time-oscillating effects of the potential (\ref{TDepenPoten2}) in exponential decays of the survival probabilities, we fit each exponential decay of the survival probability $P(t)$ to an exponential function $\gamma\exp(-t/\tau)$ with the fitting parameters $\gamma$ and $\tau$, and introduce the decay time $\tau$ as its fitting value. 
   Figure \ref{Fig2cDecayRate5d0} is a graph of such a decay time $\tau$ as a function of the parameter $b$ for the case of $a=5$.  
   This figure suggests that the decay time $\tau$ is a decreasing function of the time-oscillation amplitude $|b|$ of the potential, and is also an even function of $b$. 
   As shown in the dashed line of this figure, the $b$-dependence of the decay time $\tau$ is well fitted to a quadratic curve $\tau = \zeta_{1}-\zeta_{2}b^{2}$ with constants $\zeta_{1} = 2.61$ and $\zeta_{2} = 8.11\times 10^{-2}$ near $b=0$, although a deviation from it is recognized for a large value of $|b|$. 
%

\begin{figure}[!t] 
\vspfigA
\begin{center}
\includegraphics[width=\widthfig]{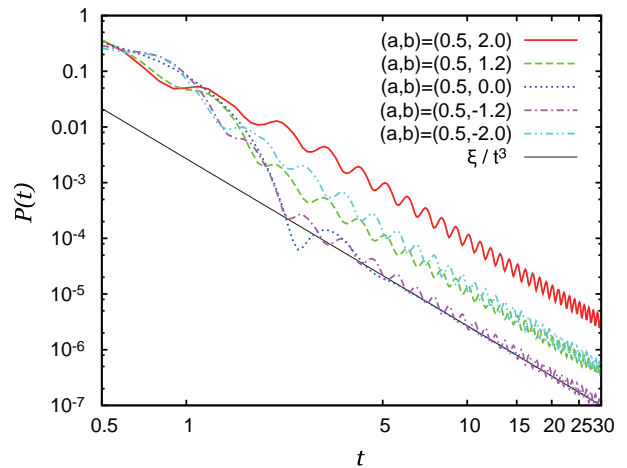}
\vspfigB
\caption{
   (Color online) 
   The survival probabilities $P(t)$ as functions of time $t$ for the cases of $(a,b) = (0.5,2)$ (the solid red line), $(a,b) = (0.5,1.2)$ (the dashed green line), and  $(a,b) = (0.5,0)$ (the dotted blue line), $(a,b) = (0.5,-1.2)$ (the dash-dotted purple line) and $(a,b) = (0.5,-2)$ (the dash-double-dotted cyan line) on log-log plots. 
   The straight shin black line shows the asymptotic power decay $\xi/t^{3}$ for the case of $b=0$. 
   }
\label{Fig2dSurProb0d5}
\end{center}
\vspfigC
\end{figure}  
%
\begin{figure}[!t] 
\vspfigA
\begin{center}
\includegraphics[width=\widthfig]{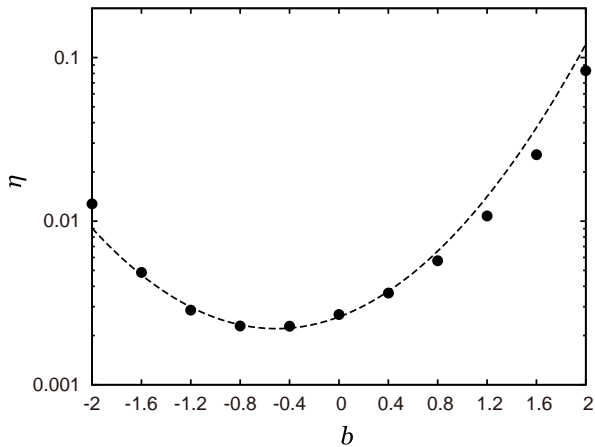}
\vspfigB
\caption{
   The power decay magnitude $\eta$ of the survival probability decaying as a power as a function of the parameter $b$ of the potential $U(x,t) = [a + b\cos(\omega t)]\delta(x-L)$ for the case of $a=0.5$ on a linear-log plot. 
   The dashed line is a fitting curve given by $\eta = \varsigma_{1} \exp\left[\varsigma_{2} (b - \varsigma_{3})^{2}\right]$ for the numerical data of $\eta$ with fitting parameters $\varsigma_{1}$, $\varsigma_{2}$ and $\varsigma_{3}$. 
   }
\label{Fig2ePowerAmp0d5}
\end{center}
\vspfigC
\end{figure}  

\subsubsection{Power decay of the survival probability} 
\label{PowerDecayPart}
   
   In Fig. \ref{Fig2dSurProb0d5} we show the graphs of the survival probabilities $P(t)$ as functions of time $t$ for the cases of $(a,b) = (0.5,2)$ (the solid red line), $(a,b) = (0.5,1.2)$ (the dashed green line),  $(a,b) = (0.5,0)$ (the dotted blue line), $(a,b) = (0.5,-1.2)$ (the dash-dotted purple line) and $(a,b) = (0.5,-2)$ (the dash-double-dotted cyan line). 
   This figure is their log-log plots to show their power decays as straight lines, and also shows the asymptotic power decay $\xi/t^{3}$ obtained analytically for the case of $b=0$.
   Here, we chose a smaller value of the parameter $a$ than that in Fig. \ref{Fig2aSurProb1d5}, so that power decay behaviors of the survival probabilities appear more clearly.

   As shown in Fig. \ref{Fig2dSurProb0d5}, after a finite time (approximately $t>5$) the survival probabilities $P(t)$ decay as a power $\propto t^{-3}$, apart from their time-oscillations.
   The survival probability in such a power decay has a tendency to take a large value for a large value of the time-oscillation amplitude $|b|$ of the potential (\ref{TDepenPoten2}), but such increasing rate of the survival probability for $b<0$ is different from that for $b>0$.    
   The time-oscillation of the potential also causes a time-oscillation of the survival probability $P(t)$, whose amplitude is larger for a larger value of $|b|$. 
   The phase of time-oscillations of the survival probabilities for $b<0$ is opposite to that for $b>0$.     
   It may be noted that the time-oscillations of the survival probabilities in their power decays involve clear temporal increases of the value of $P(t)$ as the time $t$ passes, differently from the ones in their exponential decays.

   In order to investigate quantitatively enhancement of the survival probabilities by the time-oscillation of the potential (\ref{TDepenPoten2}) in their power decays, we fit them to the power decay function $\eta t^{-3}$ with the fitting parameter $\eta$ as the power decay magnitude, apart from their time-oscillations.
   Figure \ref{Fig2ePowerAmp0d5} is a graph of such a fitting value of $\eta$ as a function of the parameter $b$ of the potential (\ref{TDepenPoten2}). 
   Here, we used the analytical value $\eta = \xi \approx 2.69\times 10^{-3}$ only for the case of $b=0$. 
   This figure clearly demonstrates a tendency in which the power decay magnitude $\eta$ of the survival probability decaying as a power takes a large value for a large time-oscillation amplitude $|b|$ of the potential (\ref{TDepenPoten2}). 
   However, it is important to note that the power decay magnitudes $\eta$ is not an even function of $b$, and it takes the minimum value at a non-zero negative value of $b$. 
   To clarify this point, we fitted  the numerical data of $\eta$ around its minimum point to the function $\eta = \varsigma_{1} \exp\left[\varsigma_{2} (b - \varsigma_{3})^{2}\right]$ with fitting parameters $\varsigma_{1}$, $\varsigma_{2}$ and $\varsigma_{3}$. 
   Here, we used a quadratic fitting function to the data of $\log\eta$ rather than of $\eta$ itself for a better fit for a wider values of $b$, and the used values of the fitting parameters are $ \varsigma_{1}=2.20\times 10^{-3}$, $\varsigma_{2}=6.36\times 10^{-1}$ and $\varsigma_{3}=-5.09\times 10^{-1}$. 
   This function fits our numerical results of  $\eta$ reasonably well around its minimum point, and suggests that the power decay magnitudes $\eta$ takes its minimum value at $b \approx -0.5$ approximately, for the case of $a=0.5$. 
%

\section{Conclusion and remarks}

   In this paper we discussed quantum escapes of a particle via a confining potential with a time-dependent magnitude in a semi-infinite one-dimensional space. 
   We calculated scattering states of the quantum open system, and based on their orthogonal relations we derived the quantum propagator for the time-evolution of wave function of the system with a time-dependent potential. 
   Our special interest in such a quantum open system was decay properties of the survival probability, i.e. the probability for a particle to remain in the finite region between the wall and the confining potential. 
   By using the propagator represented by the scattering states, we calculated the survival probability for the case of a delta-function confining potential with a time-oscillating magnitude, and the initial state as the energy eigenstate of a particle confined in the finite region. 
   The survival probability shows two different types of decay behaviors: an exponential decay at early times and a power decay later. 
   The time-oscillation of the potential magnitude reduces a decay time of the survival probability decaying exponentially. 
   In contrast, a large time-oscillation of the confining potential enhances a magnitude of the survival probability decaying as a power. 
   The power decay magnitude of the survival probability is not symmetric under change of the sign of the time-oscillating part of the potential, different from the decay time of the survival probability decaying exponentially.

   In this paper, we calculated numerically the double integrals over wave numbers for the survival probability. 
   It may be noted that such numerical integrations for a time-dependent potential are much harder than the one for the corresponding time-independent potential because of more complicate time-dependence of the propagator and matrix treatments. 
   This kind of difficulties required us some additional numerical techniques (such as a calculation of the  matrix $\Lambda_{\beta}(I-\Lambda_{\beta})^{-1}$ as solutions of linear algebraic equations, etc.), and actually restricted us parameter values under which we could calculate the survival probability within a reasonable calculation time and a numerical accuracy. 
   It would be a future problem to improve further numerical calculation methods, as well as analytical ones, to calculate the survival probability for wide range of values of various parameters.

\section*{Acknowledgements}

   One of the authors (T.T.) is grateful to T. Okamura for many valuable comments and suggestions on scattering states and decay behaviors in quantum open systems.

\appendix
\section{Lippmann-Schwinger equation in the semi-infinite one-dimensional space}
\label{GreenFunctionHardWall}

   In this appendix, we explain the Lippmann-Schwinger equation (\ref{LippmSchwi1}) with the outgoing  Green function (\ref{GreenFunct1}) as the solution of the differential equation (\ref{GreenDiffe1}).  

   We note that the function $\sin(kx)$ of $k$ and $x(>0)$, proportional to the energy eigenstate with a wave number $k$ at position $x$ for the case of no potential in the semi-infinite one-dimensional space $(0,+\infty)$, satisfies the relation 
\begin{eqnarray}
   \delta (x-y) = \frac{1}{\pi}\int_{-\infty}^{+\infty} dk\; \sin(kx)\sin(ky)
\label{DeltaFunct1}
\end{eqnarray}
for any real numbers $x$ and $y$ satisfying the condition $xy > 0$. 
By Eq. (\ref{DeltaFunct1}) the Green function $G_{k}^{(n)}(x,y)$ satisfying Eq. (\ref{GreenDiffe1}), corresponding to outgoing scattering states, is given by 
\begin{widetext}
\begin{eqnarray}
   G_{k}^{(n)}(x,y) &=& \lim_{\epsilon\rightarrow+0}\left(\frac{\partial^{2} }{\partial x^{2}} 
       + \Upsilon_{k}^{(n)} + i\epsilon \right)^{-1}\delta (x-y) 
      \nonumber \\
   &=& \lim_{\epsilon\rightarrow+0}\left(\frac{\partial^{2} }{\partial x^{2}} 
      + \Upsilon_{k}^{(n)} + i\epsilon \right)^{-1}
      \frac{1}{\pi}\int_{-\infty}^{+\infty} dk\; \sin(kx)\sin(ky)
      \nonumber \\
   &=& \frac{1}{4\pi}\lim_{\epsilon\rightarrow+0}\int_{-\infty}^{+\infty} dk' \; 
      \frac{e^{ik'(x+y)}+e^{-ik'(x+y)}-e^{ik'(x-y)}-e^{-ik'(x-y)}}{k'^{2} 
      - \Upsilon_{k}^{(n)} - i\epsilon } .
\label{GreenFunct2}
\end{eqnarray}
\end{widetext}
We further note the integral formula
\begin{eqnarray}
   \lim_{\epsilon\rightarrow+0} \int_{-\infty}^{+\infty} dz\; 
      \frac{e^{iz X}}{z^{2} - \Upsilon_{k}^{(n)} - i\epsilon} 
      =   \frac{\pi i}{\sqrt{\Upsilon_{k}^{(n)}}} e^{i\sqrt{\Upsilon_{k}^{(n)}}|X|}
\label{IntegFormu1}
\end{eqnarray}
where we define the square root of $\Upsilon_{k}^{(n)}$ as $\sqrt{\Upsilon_{k}^{(n)}} \equiv i\sqrt{|\Upsilon_{k}^{(n)}|}$ for the case of $\Upsilon_{k}^{(n)}<0$. 
By carrying out the integral in Eq. (\ref{GreenFunct2}) using the formula (\ref{IntegFormu1}) we obtain Eq. (\ref{GreenFunct1}). 

For any function $\varrho_{0}(x)$ satisfying the equation $d^{2} \varrho_{0}(x)/dx^{2} + \Upsilon_{k}^{(n)} \varrho_{0}(x) = 0$, the function $\phi_{k}^{(n)}(x) = \varrho_{0}(x) + \frac{2m}{\hbar^2}\sum_{\nu =-\infty}^{+\infty} \int_{0}^{+\infty} dy\; G_{k}^{(n)}(x,y) u^{(\nu-n)}(y)  \phi_{k}^{(\nu)}(y)$ satisfies Eq. (\ref{SchruEquat1}). 
   By taking $\varrho_{0}(x) = \delta_{n0} \sqrt{\frac{2}{\pi}}\sin(kx)$ in this function $\phi_{k}^{(n)}(x)$ as a special case, we obtain Eq. (\ref{LippmSchwi1}).

\section{Orthogonal relation of scattering states with a time-dependent potential}
\label{OrthogonalRelation}

   In this appendix, we give a derivation of Eq. (\ref{OrthoRelat1}) on the orthogonalization of the scattering state $\Phi_{k}(x,t)$ of the system with the time-dependent potential given by Eqs.  (\ref{TDepenPoten1}) and (\ref{DeltaPoten1}). 
   
   
   As preparations of later calculations we note the mathematical identity
\begin{eqnarray}
   \int_{0}^{+\infty} dx\; e^{i\kappa x} 
   &\equiv& \lim_{\epsilon\rightarrow +0} \int_{0}^{+\infty} dx\; e^{i\kappa x-\epsilon x}
   \nonumber \\
   &=& \pi d (\kappa) +i \chi(\kappa) .
\label{MatheIdent1}
\end{eqnarray}
for any complex number $\kappa$ whose imaginary part is non-negative. 
   Here, $d (\kappa)$ and $\chi(\kappa)$ are defined by 
\begin{eqnarray}
   d (\kappa) &\equiv& \lim_{\epsilon\rightarrow +0} 
      \frac{1}{\pi}\frac{\epsilon}{\kappa^{2}+\epsilon^{2}} , 
      \label{FunctDel1} \\
   \chi(\kappa) &\equiv& \lim_{\epsilon\rightarrow +0} \frac{\kappa}{\kappa^{2}+\epsilon^{2}} ,
      \label{FunctChi1}
\end{eqnarray}
respectively. 
   The function (\ref{FunctDel1}) is an even function of $\kappa$, while the function (\ref{FunctChi1}) is an odd function of $\kappa$. 
   It is also noted that the function $d (\kappa)$ becomes the delta function $d (\kappa) = \delta (k)$ for any real number $\kappa=k$. 
We also note  
\begin{eqnarray}
   \int_{0}^{L} dx\; e^{i\kappa x} 
   = \left(1-e^{i\kappa L}\right) i \chi(\kappa) ,
\label{MatheIdent2}
\end{eqnarray}
for any complex number $\kappa$.  
   
   We note the relation 
\begin{eqnarray}
   \int_{0}^{+\infty} dx \; \sin(\alpha x)\sin(\beta x) 
      = \frac{\pi}{2}\left[d (\alpha -\beta) - d (\alpha +\beta) \right]
      \nonumber \\
\label{OrthoBasic1}
\end{eqnarray}
for any real numbers $\alpha$ and $\beta$. 
   Moreover, by using Eqs. (\ref{MatheIdent1}) and (\ref{MatheIdent2}) we can show
\begin{widetext}
\begin{eqnarray}
   && \int_{0}^{+\infty} dx \; \sin(\alpha x) G_{\beta}^{(\nu)}(x,L)
      \nonumber \\
   &&\spaEq 
      =  i \frac{\pi}{2}
       e^{-i\sqrt{\Upsilon_{\beta}^{(\nu)}}L} G_{\beta}^{(\nu)}(L,L)
      \left[ d\!\left(\alpha-\sqrt{\Upsilon_{\beta}^{(\nu)}}\right) 
      - d\!\left(\alpha+\sqrt{\Upsilon_{\beta}^{(\nu)}}\right) \right]
      -\sin(\alpha L) \; W\!\left(\alpha,\sqrt{\Upsilon_{\beta}^{(\nu)}}\right)  ,
      \;\;\;
\label{OrthoBasic2}
\end{eqnarray}
\begin{eqnarray}
    && \int_{0}^{+\infty} dx \;  \left[G_{\alpha}^{(\mu)}(x,L)\right]^{*} G_{\beta}^{(\nu)}(x,L) 
       \nonumber \\
    &&\spaEq  = 
       \pi e^{i\left[\left(\sqrt{\Upsilon_{\alpha}^{(\mu)}}\right)^{*}
          -\sqrt{\Upsilon_{\beta}^{(\nu)}}\right]L}
       \left[G_{\alpha}^{(\mu)}(L,L)\right]^{*}G_{\beta}^{(\nu)}(L,L)
       d\!\left(\left(\sqrt{\Upsilon_{\alpha}^{(\mu)}}\right)^{*}
      -\sqrt{\Upsilon_{\beta}^{(\nu)}}  \right)  
      \nonumber \\
   &&\spaEq\spaEq
      - \left[G_{\alpha}^{(\mu)}(L,L)\right]^{*}
      W\!\left(\left(\sqrt{\Upsilon_{\alpha}^{(\mu)}}\right)^{*}
      ,\sqrt{\Upsilon_{\beta}^{(\nu)}}\right)
      - G_{\beta}^{(\nu)}(L,L)
      W\!\left(\sqrt{\Upsilon_{\beta}^{(\nu)}}
         ,\left(\sqrt{\Upsilon_{\alpha}^{(\mu)}}\right)^{*}\right) . 
      \;\;\;
\label{OrthoBasic3}
\end{eqnarray}
\end{widetext}
Here, we used the function $W(X,Y)$ of $X$ and $Y$ defined by 
\begin{eqnarray}
   W(X,Y) &\equiv& \frac{\chi(X-Y) - \chi(X+Y)}{2Y} ,
\label{FunctW1}
\end{eqnarray}
so that we obtain 
\begin{eqnarray}
   \frac{\chi(X-Y) + \chi(X+Y)}{2X} = - W(Y,X)
\label{FunctW2}
\end{eqnarray}
using the relation $\chi(X-Y) = - \chi(Y-X)$.

   By Eq. (\ref{ScattState1}) the inner product $\int_{0}^{+\infty} dx \; \left[\Phi_{\alpha}(x,t)\right]^{*}\Phi_{\beta}(x,t) $ of the scattering states is represented as 
\begin{widetext}
\begin{eqnarray}
   \int_{0}^{+\infty} dx \; \left[\Phi_{\alpha}(x,t)\right]^{*}\Phi_{\beta}(x,t) 
   = \sum_{\mu=-\infty}^{+\infty} \sum_{\nu=-\infty}^{+\infty} 
   e^{i[E_{\alpha}-E_{\beta}+(\mu-\nu)\hbar \omega]t/\hbar} 
   \int_{0}^{+\infty} dx \; \left[\phi_{\alpha}^{(\mu)}(x)\right]^{*}\phi_{\beta}^{(\nu)}(x) .
\label{InnerCalcu1}
\end{eqnarray}
  Further, by Eqs. (\ref{LippmSchwi3}), (\ref{OrthoBasic1}), (\ref{OrthoBasic2}) and (\ref{OrthoBasic3}) the inner product $\int_{0}^{+\infty} dx \; \left[\phi_{\alpha}^{(\mu)}(x)\right]^{*}\phi_{\beta}^{(\nu)}(x)$ in the right-hand side of Eq. (\ref{InnerCalcu1}) is given by
\begin{eqnarray}
   && \int_{0}^{+\infty} dx \; \left[\phi_{\alpha}^{(\mu)}(x)\right]^{*}\phi_{\beta}^{(\nu)}(x)
      \nonumber \\
   &&\spaEq=  \delta_{\mu 0}\delta_{\nu 0} \frac{2}{\pi}
      \int_{0}^{+\infty} dx \;  \sin(\alpha x)\sin(\beta x) 
      \nonumber \\
   &&\spaEq\spaEq + \delta_{\mu 0} 
      \frac{\left[\Lambda_{\beta}(I-\Lambda_{\beta})^{-1} \right]^{(\nu 0)}\sin(\beta L)}
      {G_{\beta}^{(\nu)}(L,L) }
      \frac{2}{\pi} \int_{0}^{+\infty} dx \;  \sin(\alpha x)G_{\beta}^{(\nu)}(x,L)
      \nonumber \\
   &&\spaEq\spaEq + \delta_{\nu 0} 
      \left\{\frac{\left[\Lambda_{\alpha}(I-\Lambda_{\alpha})^{-1} \right]^{(\mu 0)}\sin(\alpha L)}
         {G_{\alpha}^{(\mu)}(L,L) }\right\}^{*}
      \frac{2}{\pi}\int_{0}^{+\infty} dx \;  \left[G_{\alpha}^{(\mu)}(x,L)\right]^{*} \sin(\beta x)
      \nonumber \\
   &&\spaEq\spaEq +  
      \left\{\frac{\left[\Lambda_{\alpha}(I-\Lambda_{\alpha})^{-1} \right]^{(\mu 0)}\sin(\alpha L)}
         {G_{\alpha}^{(\mu)}(L,L) }\right\}^{*}
      \frac{\left[\Lambda_{\beta}(I-\Lambda_{\beta})^{-1} \right]^{(\nu 0)}\sin(\beta L)}
         {G_{\beta}^{(\nu)}(L,L) }
      \frac{2}{\pi}\int_{0}^{+\infty} dx \;  
      \left[G_{\alpha}^{(\mu)}(x,L)\right]^{*} G_{\beta}^{(\nu)}(x,L) 
      \nonumber \\
   &&\spaEq 
      =  \delta_{\mu 0}\delta_{\nu 0} \left[ d (\alpha -\beta) - d (\alpha +\beta)\right]
       +\sin(\alpha L)\sin(\beta L)\left(\caA_{(\alpha,\beta)}^{(\mu,\nu)}
       + \frac{2}{\pi} \caB_{(\alpha,\beta)}^{(\mu,\nu)}\right)
\label{InnerCalcu2}
\end{eqnarray}
where $\caA_{(\alpha,\beta)}^{(\mu,\nu)}$ and  $\caB_{(\alpha,\beta)}^{(\mu,\nu)}$ are defined by 
\begin{eqnarray}
   \caA_{(\alpha,\beta)}^{(\mu,\nu)}
      &\equiv& 
      2 e^{i\left[\left(\sqrt{\Upsilon_{\alpha}^{(\mu)}}\right)^{*}
         -\sqrt{\Upsilon_{\beta}^{(\nu)}}\right]L}
      \left\{\left[\Lambda_{\alpha}(I-\Lambda_{\alpha})^{-1} \right]^{(\mu 0)}\right\}^{*}
      \left[\Lambda_{\beta}(I-\Lambda_{\beta})^{-1} \right]^{(\nu 0)}
      d\!\left(\left(\sqrt{\Upsilon_{\alpha}^{(\mu)}}\right)^{*}
      -\sqrt{\Upsilon_{\beta}^{(\nu)}}  \right) 
      \nonumber \\
   &&\spaEq 
      + i \delta_{\mu 0} \frac{e^{-i\sqrt{\Upsilon_{\beta}^{(\nu)}}L}}{\sin(\alpha L)} 
      \left[\Lambda_{\beta}(I-\Lambda_{\beta})^{-1} \right]^{(\nu 0)} 
      \left[d\!\left(\alpha-\sqrt{\Upsilon_{\beta}^{(\nu)}}\right) 
      - d\!\left(\alpha+\sqrt{\Upsilon_{\beta}^{(\nu)}}\right) \right]
      \nonumber \\
   &&\spaEq 
      - i\delta_{\nu 0} \frac{e^{i\left(\sqrt{\Upsilon_{\alpha}^{(\mu)}}\right)^{*}L}}{\sin(\beta L)}
      \left\{\left[\Lambda_{\alpha}(I-\Lambda_{\alpha})^{-1} \right]^{(\mu 0)}   \right\}^{*}
      \left[d\!\left(\beta-\left(\sqrt{\Upsilon_{\alpha}^{(\mu)}}\right)^{*}\right) 
      - d\!\left(\beta+\left(\sqrt{\Upsilon_{\alpha}^{(\mu)}}\right)^{*}\right) \right]  ,
      \nonumber \\
\label{FunctA1} \\
   \caB_{(\alpha,\beta)}^{(\mu,\nu)}
      &\equiv& -\left\{\left[\Lambda_{\alpha}(I-\Lambda_{\alpha})^{-1} \right]^{(\mu 0)}
      \right\}^{*}
      \frac{\left[\Lambda_{\beta}(I-\Lambda_{\beta})^{-1} \right]^{(\nu 0)}}
         {G_{\beta}^{(\nu)}(L,L) }
      W\!\left(\left(\sqrt{\Upsilon_{\alpha}^{(\mu)}}\right)^{*}
      ,\sqrt{\Upsilon_{\beta}^{(\nu)}}\right)
      \nonumber \\
   &&\spaEq
      - \left\{\frac{\left[\Lambda_{\alpha}(I-\Lambda_{\alpha})^{-1} \right]^{(\mu 0)}}
         {G_{\alpha}^{(\mu)}(L,L) }\right\}^{*}
     \left[\Lambda_{\beta}(I-\Lambda_{\beta})^{-1} \right]^{(\nu 0)}
     W\!\left(\sqrt{\Upsilon_{\beta}^{(\nu)}}
         ,\left(\sqrt{\Upsilon_{\alpha}^{(\mu)}}\right)^{*}\right) 
      \nonumber \\
   &&\spaEq  
      -\delta_{\mu 0} \frac{\left[\Lambda_{\beta}(I-\Lambda_{\beta})^{-1} \right]^{(\nu 0)}}
         {G_{\beta}^{(\nu)}(L,L) }
      W\!\left(\alpha,\sqrt{\Upsilon_{\beta}^{(\nu)}}\right) 
     - \delta_{\nu 0} 
      \left\{\frac{\left[\Lambda_{\alpha}(I-\Lambda_{\alpha})^{-1} \right]^{(\mu 0)}}
         {G_{\alpha}^{(\mu)}(L,L) }\right\}^{*}
       W\!\left(\beta,\left(\sqrt{\Upsilon_{\alpha}^{(\mu)}}\right)^{*}\right)  .
       \nonumber \\
\label{FunctB1}
\end{eqnarray}
\end{widetext}
Here, the quantity $\caA_{(\alpha,\beta)}^{(\mu,\nu)}$ includes the type of $d\!\left(\left(\sqrt{\Upsilon_{\alpha}^{(\mu)}}\right)^{*} \pm \sqrt{\Upsilon_{\beta}^{(\nu)}}\right) $ and the quantity  $\caB_{(\alpha,\beta)}^{(\mu,\nu)}$ includes the terms with the function $W(x,y)$.

   Now, we note that for the function (\ref{FunctDel1}) we have 
\begin{eqnarray}
   X(\alpha)d(\alpha-\beta) = X(\beta)d(\alpha-\beta)
\label{DFunctIdent1}
\end{eqnarray}
for any function $X(\kappa)$ of complex number $\kappa$ 
because of $d(\alpha-\beta) = 0$ for any $\alpha\neq\beta$. 
   Noting Eq. (\ref{DFunctIdent1}) we obtain
\begin{widetext}
\begin{eqnarray}
   && G_{\beta}^{(\nu)}(L,L) d\!\left(\alpha \pm \sqrt{\Upsilon_{\beta}^{(\nu)}}  \right)
   = \pm \frac{e^{i\sqrt{\Upsilon_{\beta}^{(\nu)}} L}\sin(\alpha L)}{\sqrt{\Upsilon_{\beta}^{(\nu)}}} 
      d\!\left(\alpha \pm\sqrt{\Upsilon_{\beta}^{(\nu)}}  \right) ,
      \label{GreenDeltaSup1} \\
   &&\left[G_{\alpha}^{(\mu)}(L,L) \right]^{*}G_{\beta}^{(\nu)}(L,L) 
   d\!\left(\left(\sqrt{\Upsilon_{\alpha}^{(\mu)}}\right)^{*}
      -\sqrt{\Upsilon_{\beta}^{(\nu)}}  \right)
      \nonumber \\
   &&\spaEq = 
      \left\{ \frac{\left[G_{\alpha}^{(\mu)}(L,L) \right]^{*}}{2i\sqrt{\Upsilon_{\beta}^{(\nu)}}}
      - \frac{G_{\beta}^{(\nu)}(L,L)}{2i\left(\sqrt{\Upsilon_{\alpha}^{(\mu)}}\right)^{*}} \right\}
      d\!\left(\left(\sqrt{\Upsilon_{\alpha}^{(\mu)}}\right)^{*}
      -\sqrt{\Upsilon_{\beta}^{(\nu)}}  \right)
      \label{GreenDeltaSup2a}  \\
   &&\spaEq = 
       -i\left\{\left[G_{\alpha}^{(\mu)}(L,L) \right]^{*} - G_{\beta}^{(\nu)}(L,L) \right\}
       D\!\left(\left(\sqrt{\Upsilon_{\alpha}^{(\mu)}}\right)^{*}
      ,\sqrt{\Upsilon_{\beta}^{(\nu)}}  \right) .
\label{GreenDeltaSup2b} 
\end{eqnarray}
Here, we used the relation 
\begin{eqnarray}
   \left\{\left[G_{\alpha}^{(\mu)}(L,L) \right]^{*} - G_{\beta}^{(\nu)}(L,L) \right\}
   d\!\left(\left(\sqrt{\Upsilon_{\alpha}^{(\mu)}}\right)^{*}
      +\sqrt{\Upsilon_{\beta}^{(\nu)}}  \right) = 0
\label{GreenDeltaIdent1}
\end{eqnarray}
\end{widetext}
to derive Eq. (\ref{GreenDeltaSup2b}) from Eq. (\ref{GreenDeltaSup2a}) by Eqs. (\ref{GreenFunct1}) and (\ref{DFunctIdent1}), and also the function $D(X,Y)$ of $X$ and $Y$ defined by 
\begin{eqnarray}
   D(X,Y) \equiv \frac{d(X-Y) + d(X+Y)}{2Y} ,
\label{FunctD1}
\end{eqnarray}
and for Eq. (\ref{GreenDeltaSup1}) we adopt the convention that any equation containing the symbols $\pm$ 
on the left- and right-hand sides, denote \emph{two} equations, one with only the upper symbol ($+$ in $\pm$) and the other with only the lower symbol ($-$ in $\pm$). 
   Moreover, we note the relation
\begin{eqnarray}
   \left[\Lambda_{\alpha}(I-\Lambda_{\alpha})^{-1} \right]^{(\mu 0)} 
   = -\delta_{\mu 0} +  \left[(I-\Lambda_{\alpha})^{-1} \right]^{(\mu 0)} .
\label{IdentAppen1}
\end{eqnarray}
   By Eqs. (\ref{GreenDeltaSup1}), (\ref{GreenDeltaSup2b}) and (\ref{IdentAppen1}) 
the quantity (\ref{FunctA1}) is rewritten as 
\begin{widetext}
\begin{eqnarray}
   \caA_{(\alpha,\beta)}^{(\mu,\nu)} 
   &=& - 2 i \left\{\left\{\left[(I-\Lambda_{\alpha})^{-1} \right]^{(\mu 0)}\right\}^{*}
      \frac{\left[\Lambda_{\beta}(I-\Lambda_{\beta})^{-1} \right]^{(\nu 0)} }
         {G_{\beta}^{(\nu)}(L,L)}
      \right.\nonumber \\
   &&\spaEq\spaEq\spaEq\left.
      - \left\{\frac{\left[\Lambda_{\alpha}(I-\Lambda_{\alpha})^{-1} \right]^{(\mu 0)}}
         {G_{\alpha}^{(\mu)}(L,L)}\right\}^{*}
      \left[(I-\Lambda_{\beta})^{-1} \right]^{(\nu 0)}\right\}
      D\!\left(\left(\sqrt{\Upsilon_{\alpha}^{(\mu)}}\right)^{*}
      ,\sqrt{\Upsilon_{\beta}^{(\nu)}}  \right)
      \nonumber \\
   &&\spaEq + 2i\delta_{\nu 0} 
      \left\{\frac{\left[\Lambda_{\alpha}(I-\Lambda_{\alpha})^{-1} \right]^{(\mu 0)}}
         {G_{\alpha}^{(\mu)}(L,L)\sqrt{\Upsilon_{\alpha}^{(\mu)}}}\right\}^{*} 
      d\!\left(\left(\sqrt{\Upsilon_{\alpha}^{(\mu)}}\right)^{*}+\beta\right)
\label{FunctBSup1}
\end{eqnarray}
for any positive numbers $\alpha$ and $\beta$. 
   Here, we used the relation $\delta_{\mu 0}\sqrt{\Upsilon_{\alpha}^{(\mu)}} 
 = \delta_{\mu 0} \alpha$ for any positive number $\alpha$.
   By Eq. (\ref{IdentAppen1}) we also rewrite the quantity (\ref{FunctB1}) as 
\begin{eqnarray}
   \caB_{(\alpha,\beta)}^{(\mu,\nu)} 
      &=& -\left\{\left[(I-\Lambda_{\alpha})^{-1} \right]^{(\mu 0)}
      \right\}^{*}
      \frac{\left[\Lambda_{\beta}(I-\Lambda_{\beta})^{-1} \right]^{(\nu 0)}}
         {G_{\beta}^{(\nu)}(L,L) }
      W\!\left(\left(\sqrt{\Upsilon_{\alpha}^{(\mu)}}\right)^{*}
      ,\sqrt{\Upsilon_{\beta}^{(\nu)}}\right)
      \nonumber \\
   &&\spaEq
      - \left\{\frac{\left[\Lambda_{\alpha}(I-\Lambda_{\alpha})^{-1} \right]^{(\mu 0)}}
         {G_{\alpha}^{(\mu)}(L,L) }\right\}^{*}
     \left[(I-\Lambda_{\beta})^{-1} \right]^{(\nu 0)}
     W\!\left(\sqrt{\Upsilon_{\beta}^{(\nu)}}
         ,\left(\sqrt{\Upsilon_{\alpha}^{(\mu)}}\right)^{*}\right) 
\label{FunctASup1}
\end{eqnarray}
for any positive numbers $\alpha$ and $\beta$. 
   By the definition of $\sqrt{\Upsilon_{\alpha}^{(\mu)}}$ the real part of $\sqrt{\Upsilon_{\alpha}^{(\mu)}}$ is non-negative, so that we have $d\!\left(\left(\sqrt{\Upsilon_{\alpha}^{(\mu)}}\right)^{*}+\beta\right) =0$, as well as $d(\alpha+\beta)=0$ and  $d(\alpha-\beta)=\delta(\alpha-\beta)$, for any positive numbers $\alpha$ and $\beta$.  
   By these facts and Eqs. (\ref{InnerCalcu2}), (\ref{FunctBSup1}) and (\ref{FunctASup1}) we obtain
\begin{eqnarray}
    && \int_{0}^{+\infty} dx \; \left[\phi_{\alpha}^{(\mu)}(x)\right]^{*}\phi_{\beta}^{(\nu)}(x)  
      \nonumber \\
   &&\spaEq 
      =  \delta_{\mu 0}\delta_{\nu 0} \delta (\alpha -\beta) 
      \nonumber \\
   &&\spaEq\spaEq 
       -\frac{2i}{\pi} \sin(\alpha L)\sin(\beta L)
          \Bigg\{\left\{\left[(I-\Lambda_{\alpha})^{-1} \right]^{(\mu 0)}\right\}^{*}
      \frac{\left[\Lambda_{\beta}(I-\Lambda_{\beta})^{-1} \right]^{(\nu 0)}}{G_{\beta}^{(\nu)}(L,L) }
      \nonumber \\
   &&\spaEq\spaEq\spaEq\spaEq\spaEq\spaEq\spaEq\spaEq\left.\times
      \left[\pi D\!\left(\left(\sqrt{\Upsilon_{\alpha}^{(\mu)}}\right)^{*}
         ,\sqrt{\Upsilon_{\beta}^{(\nu)}}  \right)
      - i W\!\left(\left(\sqrt{\Upsilon_{\alpha}^{(\mu)}}\right)^{*}
         ,\sqrt{\Upsilon_{\beta}^{(\nu)}}\right) \right]
      \right.\nonumber \\
   &&\spaEq\spaEq\spaEq\spaEq\left.
      - \left\{\frac{\left[\Lambda_{\alpha}(I-\Lambda_{\alpha})^{-1} \right]^{(\mu 0)}}
         {G_{\alpha}^{(\mu)}(L,L) }\right\}^{*}
     \left[(I-\Lambda_{\beta})^{-1} \right]^{(\nu 0)}
      \right.\nonumber \\
   &&\spaEq\spaEq\spaEq\spaEq\spaEq\spaEq\times
     \left[\pi D\!\left(\left(\sqrt{\Upsilon_{\alpha}^{(\mu)}}\right)^{*}
         ,\sqrt{\Upsilon_{\beta}^{(\nu)}}  \right)
      +iW\!\left(\sqrt{\Upsilon_{\beta}^{(\nu)}}
         ,\left(\sqrt{\Upsilon_{\alpha}^{(\mu)}}\right)^{*}\right) \right]\Bigg\}
\label{InnerCalcu3}
\end{eqnarray}
\end{widetext}
for any positive numbers $\alpha$ and $\beta$.

For the quantity $D\!\left(\left(\sqrt{\Upsilon_{\alpha}^{(\mu)}}\right)^{*},\sqrt{\Upsilon_{\beta}^{(\nu)}}  \right)$ we have  

\begin{eqnarray}
   && \spaNeg 
      D\!\left(\left(\sqrt{\Upsilon_{\alpha}^{(\mu)}}\right)^{*}
         ,\sqrt{\Upsilon_{\beta}^{(\nu)}}  \right)
      \nonumber \\
   &&
      =    \left\{\begin{array}{ll}
      D\!\left(\sqrt{\left|\Upsilon_{\alpha}^{(\mu)}\right|}
         ,\sqrt{\left|\Upsilon_{\beta}^{(\nu)}\right|}  \right) 
         & \mbox{for}\;\;\;  \Upsilon_{\alpha}^{(\mu)} \Upsilon_{\beta}^{(\nu)}>0 \\
      0 &  \mbox{for}\;\;\; \Upsilon_{\alpha}^{(\mu)} \Upsilon_{\beta}^{(\nu)}<0      
      \end{array}\right.
      \nonumber \\
   && 
      = \delta\!\left(\Upsilon_{\alpha}^{(\mu)} -\Upsilon_{\beta}^{(\nu)} \right)
      \nonumber \\
   && 
      = \delta\!\left(\alpha^{2} -\beta^{2} +\frac{2m\omega}{\hbar} (\mu-\nu)   \right) ,
\label{FunctDPrope1}
\end{eqnarray}
where we used Eqs. (\ref{Upsi1}), (\ref{FunctDel1})  and (\ref{FunctD1}), the relation $D(-ix, iy) = D(x, y)$, $d(x) = \delta (x)$, $d(x^{2}-y^{2}) = [\delta (x-|y|) +\delta(x+|y|)]/(2|y|)$ and $\delta (-x) = \delta (x)$ for any real number $x$ and $y$ $(\neq 0)$. 
   By Eqs. (\ref{Upsi1}), (\ref{FunctChi1}) and (\ref{FunctW1}) we further note
\begin{eqnarray}
   && W\!\left(\left(\sqrt{\Upsilon_{\alpha}^{(\mu)}}\right)^{*}
      ,\sqrt{\Upsilon_{\beta}^{(\nu)}}\right) 
      \nonumber \\
   &&\spaEq 
      = - W\!\left(\sqrt{\Upsilon_{\beta}^{(\nu)}} , 
         \left(\sqrt{\Upsilon_{\alpha}^{(\mu)}}\right)^{*} \right)
         \nonumber \\
   &&\spaEq 
      =  \chi\!\left(\Upsilon_{\alpha}^{(\mu)}-\Upsilon_{\beta}^{(\nu)}\right)
      \nonumber \\
   &&\spaEq 
      = \chi\!\left(\alpha^{2} -\beta^{2} +\frac{2m\omega}{\hbar} (\mu-\nu)   \right) 
\label{FunctWPrope1}
\end{eqnarray}
for any numbers $\alpha$ and $\beta$ except in the zero-measure region satisfying the condition $\left(\sqrt{\Upsilon_{\alpha}^{(\mu)}}\right)^{*} = \pm\sqrt{\Upsilon_{\beta}^{(\nu)}} \neq 0$, i.e. for almost all numbers $\alpha$ and $\beta$. 
   By introducing the quantity $C_{\alpha\beta}^{(n)}(t)$ as
\begin{widetext}
\begin{eqnarray}
   C_{\alpha\beta}^{(n)}(t) 
   \equiv e^{i(\alpha^{2}-\beta^{2}+2m \omega n/\hbar)\hbar^{2} t/(2m)}\left[\pi \delta\!\left(\alpha^{2} -\beta^{2} +\frac{2m\omega}{\hbar} n   \right) 
   - i \chi\!\left(\alpha^{2} -\beta^{2} +\frac{2m\omega}{\hbar} n   \right) \right] ,
\label{FunctC1}
\end{eqnarray}
and by Eqs.  (\ref{FunctLambda1}), (\ref{InnerCalcu1}), (\ref{InnerCalcu3}), (\ref{FunctDPrope1}),  (\ref{FunctWPrope1}) and $\left[\lambda^{(n)}\right]^{*} = \lambda^{(-n)}$, we obtain 
\begin{eqnarray}
   && \int_{0}^{+\infty} dx \; \left[\Phi_{\alpha}(x,t)\right]^{*}\Phi_{\beta}(x,t) 
      \nonumber \\
   &&\spaEq = \delta (\alpha -\beta) -\frac{4m i}{\pi \hbar^{2}} \sin(\alpha L)\sin(\beta L)
      \sum_{n=-\infty}^{+\infty}\left\{\left[(I-\Lambda_{\alpha})^{-1} \right]^{(n 0)}\right\}^{*}
      \nonumber \\
   &&\spaEq\spaEq\spaEq\spaEq\spaEq\spaEq\times 
        \sum_{\mu=-\infty}^{+\infty} \sum_{\nu=-\infty}^{+\infty}
          \Bigg\{
      C_{\alpha\beta}^{(n-\mu)}(t)
       \lambda^{(-(\mu-\nu))}\left[(I-\Lambda_{\beta})^{-1} \right]^{(\nu 0)}
      \nonumber \\
   &&\spaEq\spaEq\spaEq\spaEq\spaEq\spaEq\spaEq\spaEq\spaEq\spaEq\spaEq
      - \lambda^{(-(n-\mu))} C_{\alpha\beta}^{(\mu-\nu)}(t)
      \left[(I-\Lambda_{\beta})^{-1} \right]^{(\nu 0)}\Bigg\}
\label{InnerCalcu4} 
\end{eqnarray}
\end{widetext}
for almost all positive numbers $\alpha$ and $\beta$.

Now, we introduce the discretized Fourier transformation $\widetilde{X}(a)$ as
\begin{eqnarray}
   \widetilde{X}(\rho) &\equiv& \sum_{n = -\infty}^{+\infty} X_{n} e^{i\rho n}, 
      \label{DiscrFouri1} \\
   X_{n} &=& \frac{1}{2\pi}\int_{-\pi}^{+\pi} d\rho \; \widetilde{X}(\rho) e^{-i\rho n}
      \label{DiscrFouri2}
\end{eqnarray}
for any function $X_{n}$ of integer $n$, noting the formula $\int_{-\pi}^{+\pi} d\rho \; \exp (i n \rho) = 2\pi \delta_{n 0}$ for any integer $n$. 
   By using this transformation and the formula  $\sum_{n = -\infty}^{+\infty}  e^{i\rho n} = 2\pi \delta (\rho)$ for any $\rho\in [-\pi, \pi]$, the discretized Fourier transformation $\widetilde{(X\ast Y)}(\rho)$ of the convolution $(X\ast Y)_{\mu} \equiv \sum_{\nu = -\infty}^{+\infty} X_{\mu-\nu}Y_{\nu}$ for any functions $X_{n}$ and $Y_{n}$ of $n$ is represented as
\begin{eqnarray}
   \widetilde{(X\ast Y)}(\rho) = \widetilde{X}(\rho)\widetilde{Y}(\rho)
\label{ConvoFouri1}
\end{eqnarray}
for $-\pi \leq \rho \leq  \pi$, as known as the convolution formula of the Fourier transformation. 
   By using this property we have 
\begin{widetext}
\begin{eqnarray}
    \sum_{\mu=-\infty}^{+\infty} \sum_{\nu=-\infty}^{+\infty}
      C_{\alpha\beta}^{(n-\mu)}(t)
       \lambda^{(-(\mu-\nu))}\left[(I-\Lambda_{\beta})^{-1} \right]^{(\nu 0)}
      = \sum_{\mu=-\infty}^{+\infty} \sum_{\nu=-\infty}^{+\infty}
      \lambda^{(-(n-\mu))} C_{\alpha\beta}^{(\mu-\nu)}(t)
      \left[(I-\Lambda_{\beta})^{-1} \right]^{(\nu 0)}
\label{FouriTransIdent1}
\end{eqnarray}
\end{widetext}
because the discretized Fourier transformation of the left-hand side of Eq. (\ref{FouriTransIdent1}) is equal to the one of the right-hand side of Eq. (\ref{FouriTransIdent1}). 
   By inserting (\ref{FouriTransIdent1}) into Eq. (\ref{InnerCalcu4}) we obtain Eq. (\ref{OrthoRelat1}) for almost all positive numbers $\alpha$ and $\beta$.

\section{Survival Probability under the initial conditions (\ref{InitiCondi1}) and (\ref{InitiCondi2}) }
\label{SurvivalProbability}

   In this appendix we give an explicit form of the survival probability $P(t)$ under the initial conditions (\ref{InitiCondi1}) and (\ref{InitiCondi2}). 

   We note the integrals 
\begin{eqnarray}
   \int_{0}^{L} dx \; \sin(\alpha x)  \sin(\beta x) 
      = \frac{L}{2}  \caS(\alpha L,\beta L), 
\label{CoeffAInt1}
\end{eqnarray}
\begin{eqnarray}
   &&  
      \int_{0}^{L} dx \; G_{\alpha}^{(n)}(x,L) \sin(\beta x) 
      \nonumber \\
   && \spaEq 
      = - \frac{L e^{i\sqrt{\Upsilon_{\alpha}^{(n)}}L}}
         {2\sqrt{\Upsilon_{\alpha}^{(n)}}}   
      \caS\!\left(\sqrt{\Upsilon_{\alpha}^{(n)}} L,\beta L\right), 
\label{CoeffAInt2}
\end{eqnarray}
\begin{widetext}
\begin{eqnarray}
   \int_{0}^{L}dx\;  \left[G_{\alpha}^{(\mu)}(x,L)\right]^{*} G_{\beta}^{(\nu)}(x,L) 
      = \frac{L e^{-i\left[\left(\sqrt{\Upsilon_{\alpha}^{(\mu)}}\right)^{*}
            -\sqrt{\Upsilon_{\beta}^{(\nu)}}\right]L} }
         {2\left(\sqrt{\Upsilon_{\alpha}^{(\mu)}}\right)^{*}\sqrt{\Upsilon_{\beta}^{(\nu)}}}
      \caS\!\left(\left(\sqrt{\Upsilon_{\alpha}^{(\mu)}}\right)^{*} L
      ,\sqrt{\Upsilon_{\beta}^{(\nu)}}L\right) 
\label{CoeffAInt3}
\end{eqnarray}
where $\caS(X,Y)$ is the function of $X$ and $Y$, defined by
\begin{eqnarray}
   \caS(X,Y) &\equiv& \frac{1}{2i}\left[\frac{e^{i(X-Y)}-e^{-i(X-Y)}}{X-Y} 
      -  \frac{e^{i(X+Y)}-e^{-i(X+Y)}}{X+Y}\right] 
      \label{FunctS1}
\end{eqnarray}
and satisfies the relation $\caS(X,Y) = \caS(Y,X)$. 
   By using Eqs. (\ref{ScattState1}), (\ref{LippmSchwi3}), (\ref{CoeffAInt1}) and (\ref{CoeffAInt2}), for the initial state given by Eqs. (\ref{InitiCondi1}) and (\ref{InitiCondi2}) the coefficient (\ref{CoeffA1}) is represented as 
\begin{eqnarray}
   A_{k} = \sqrt{\frac{L}{\pi}}\left[\caS\!\left(kL,\sigma\pi\right)+\sum_{n=-\infty}^{+\infty} 
   \caR\!\left(k,n;0\right)\caS\!\left(L\sqrt{\Upsilon_{k}^{(n)}},\sigma\pi\right)\right]^{*} ,
\label{CoeffA2}
\end{eqnarray}
where we define the function $\caR(k,n;t)$ of $k$, $n$ and $t$ by 
\begin{eqnarray}
   \caR(k,n;t) 
   &\equiv& -\frac{\left[\Lambda_{k}(I-\Lambda_{k})^{-1} \right]^{(n0)}\sin(kL)}{
      G_{k}^{(n)}(L,L)\sqrt{\Upsilon_{k}^{(n)}}}
      e^{i \left(\sqrt{\Upsilon_{k}^{(n)}}L-n\omega t\right)} .
      \label{FunctR1}
\end{eqnarray}
\end{widetext}
By Eqs. (\ref{ScattState1}), (\ref{LippmSchwi3}), (\ref{WaveFunct1}), (\ref{CoeffAInt1}), (\ref{CoeffAInt2}), (\ref{CoeffAInt3}), (\ref{CoeffA2}) and (\ref{FunctR1}), the survival probability (\ref{SurviProba1}) is represented as 
\begin{eqnarray}
   P(t) = \int_{0}^{+\infty}d\alpha\; \int_{0}^{+\infty}d\beta\; 
      \left( A_{\alpha}\right)^{*}A_{\beta} \; B_{\alpha\beta}(t)
\label{SurviProba2}
\end{eqnarray}
where $B_{\alpha\beta}(t)$ is defined by 
\begin{widetext}
\begin{eqnarray}
   B_{\alpha\beta} &\equiv&  \frac{L}{\pi} 
      \Biggl\{\caS(\alpha L,\beta L)
      + \sum_{\mu=-\infty}^{+\infty} 
      \left[\caR(\alpha,\mu;t) \caS\!\left(\sqrt{\Upsilon_{\alpha}^{(\mu)}}L,\beta L\right)\right]^{*}
      +\sum_{\nu=-\infty}^{+\infty} 
      \caR(\beta,\nu;t) \caS\!\left(\alpha L,\sqrt{\Upsilon_{\beta}^{(\nu)}}L\right)
      \nonumber \\
   &&\spaEq 
      + \sum_{\mu=-\infty}^{+\infty} \sum_{\nu=-\infty}^{+\infty} 
      \left[\caR(\alpha,\mu;t) \right]^{*}
      \caR(\beta,\nu;t) \caS\!\left(\left(\sqrt{\Upsilon_{\alpha}^{(\mu)}}\right)^{*} L
      ,\sqrt{\Upsilon_{\beta}^{(\nu)}}L\right) \Biggr\}
      e^{i(E_{\alpha}-E_{\beta})t/\hbar}  .
\label{FunctBSurPro1}
\end{eqnarray}
\end{widetext}
   Therefore, the calculation of the survival probability $P(t)$ is attributed into the calculations of the quantities $A_{k}$ and $B_{\alpha\beta}(t)$, and the double integrals (\ref{SurviProba2}). 



%
%


\begin{thebibliography}{99}
\bibitem{Memo2a} Our numerical calculations confirm the condition $P(0)=1$ for the initial conditions  (\ref{InitiCondi1}) and (\ref{InitiCondi2}). 
   This is a supportive evidence that the initial state described by Eqs. (\ref{InitiCondi1}) and (\ref{InitiCondi2}) is expanded by the scattering states $\{\Phi_{k}(x,t)\}_{k}$, namely the condition (\ref{WaveFunct1}).
\end{thebibliography}
\end{document}